\documentclass{PoS}

\title{Calculating Quark Number Susceptibilities with Domain-Wall Fermions}

\ShortTitle{Calculating Quark Number Susceptibilities with Domain-Wall Fermions}

\author{\speaker{\bf Prasad Hegde}$^{,a}$,  {\bf Frithjof Karsch}$^{b,c}$, 
         {\bf Christian Schmidt}$^c$ {\bf (RBC Collaboration)}\\ \\
        $^a$ Dept. of Physics and Astronomy, SUNY Stony Brook, Stony Brook, NY 11790, USA.\\
        ~~~E-mail: \email{phegde@quark.phy.bnl.gov}\\ 
	$^b$ Department of Physics, Brookhaven National Laboratories, Upton, NY 11733, USA.\\
	~~~E-mail: \email{karsch@quark.phy.bnl.gov}\\ 
	$^c$ Fakult\"at f\"ur Physik, Universit\"at Bielefeld, D-33615 Bielefeld, Germany.\\
	~~~E-mail: \email{schmidt@physik.uni-bielefeld.de}}

\abstract{We present results from calculations of different quark
number and hadronic susceptibilities on 2+1-flavor dynamical domain wall
ensembles. We find that the iso-spin and electric charge susceptibilities 
are especially well suited to determine the transition temperature, as these
quantities show only small statistical errors. Moreover, the transition values of the
coupling obtained from iso-spin and electrical charge susceptibilities are in
good agreement with the one obtained from the chiral condensate.}

\FullConference{The XXVI International Symposium on Lattice Field Theory\\
		 July 14-19 2008\\
		 Williamsburg, Virginia, USA}

\usepackage{amsmath}
\newcommand{\cu}{c_2^{u}}
\newcommand{\cs}{c_2^{s}}
\newcommand{\cI}{c_2^{I}}
\newcommand{\cQ}{c_2^{Q}}
\newcommand{\cud}{c_{11}^{ud}}
\newcommand{\cus}{c_{11}^{us}}
\newcommand{\Minv}{M^{-1}}
\renewcommand{\d}{\mathrm{d}}
\newcommand{\dM}[1]{\frac{\d M}{\d \mu_{#1}}}
\newcommand{\ddM}[1]{\frac{\d^2 M}{\d \mu^2_{#1}}}
\newcommand{\tr}{\text{tr}}

\begin{document}
\section{Introduction}
Sufficiently hot and dense hadronic matter undergoes a transition
from a confined hadronic phase 
to a deconfined, chirally symmetric medium, the quark gluon plasma
(QGP). At zero chemical potential, the transition has been established
by lattice QCD simulations, which have proven to be a powerful
method to analyze the non-perturbative features of the QGP close to
the transition region.  At non-zero chemical potential, however,
lattice simulations are limited by the sign problem. In fact, direct
simulations by standard Monte Carlo methods are not possible. In order
to evaluate the response of the medium to a small non-zero chemical
potential, we perform a Taylor expansion of the pressure (grand
canonical potential) \cite{Allton:2002zi}. For the dimensionless
combination $p/T^4$ we define the expansion coefficients $c_{ijk}$ as
\begin{equation}
\frac{p}{T^4}-\left.\frac{p}{T^4}\right|_{\mu=0} = \sum_{i,j,k} 
c_{ijk}\left(\frac{\mu_u}{T}\right)^i \left(\frac{\mu_d}{T}\right)^j
\left(\frac{\mu_s}{T}\right)^k.
\label{eq:c_def}
\end{equation}
Here $\mu_{u,d,s}$ are the chemical potentials of up-, down- and
strange-quarks, respectively.  The coefficients $c_{i,j,k}$ have been
calculated previously using staggered
\cite{six_order,cschmidt_and_cmiao,gavai_gupta,milc} and Wilson 
\cite{whot} quarks.  As they are evaluated at vanishing chemical potential
they also provide important information about thermodynamic properties of QCD
at vanishing baryon number density. 
%In particular they reflect the nature
%of relevant degrees of freedom at a given temperature and quantify the 
%temperature dependence of fluctuations and correlations among various 
%charges, e.g.  baryon number, electric charge, strangeness as well as iso-spin. 

We present here preliminary results on the smallest
non-vanishing coefficients\footnote{It can be shown that the odd-order
coefficients vanish through CP-symmetry.} which are $c_{200} \equiv
\cu$, $c_{002} \equiv \cs$, $c_{110} \equiv \cud$ and $c_{101} \equiv
\cus$.\footnote{In the 2+1-flavor theory up and down quark masses are
degenerate. The coefficients $c_{ijk}$
and $c_{jik}$ are thus equal.}  Apart from these basic coefficients
characterizing fluctuations and correlations for different quark
flavors, we consider combinations of them that yield
iso-spin (I) and electric charge (Q) fluctuations,  
\begin{align}
\cI &= \frac{1}{4}\left(2 \cu - \cud\right), \notag \\
\cQ &= \frac{1}{9}\left(5 \cu + \cs - 2 \cud - \cus\right).
\label{eq:cQI_def}
\end{align}

In general, the second order coefficients are related to 
the quadratic fluctuations, also known as susceptibilities  $\chi_2^X$,
of the corresponding charge densities ($n_X$). Fourth order coefficients 
are related to the quartic fluctuations $\chi_4$. In 
terms of the expansion coefficients we have
\begin{eqnarray}
\frac{\chi_2^{X}}{T^2}&\equiv&\langle n_X^2 \rangle=2 c_2^{X},
\nonumber\\
\frac{\chi_4^{X}}{T^2}&\equiv&\langle n_X^4 \rangle 
- 3\langle n_X^2 \rangle^2=24 c_4^{X} 
\qquad\mbox{with}\qquad X\in\{u,d,s,I,Q,\dots\}. 
\end{eqnarray}

\begin{figure}[!tbh]
\centering
\includegraphics[width=0.49\textwidth,height=0.2\textheight]{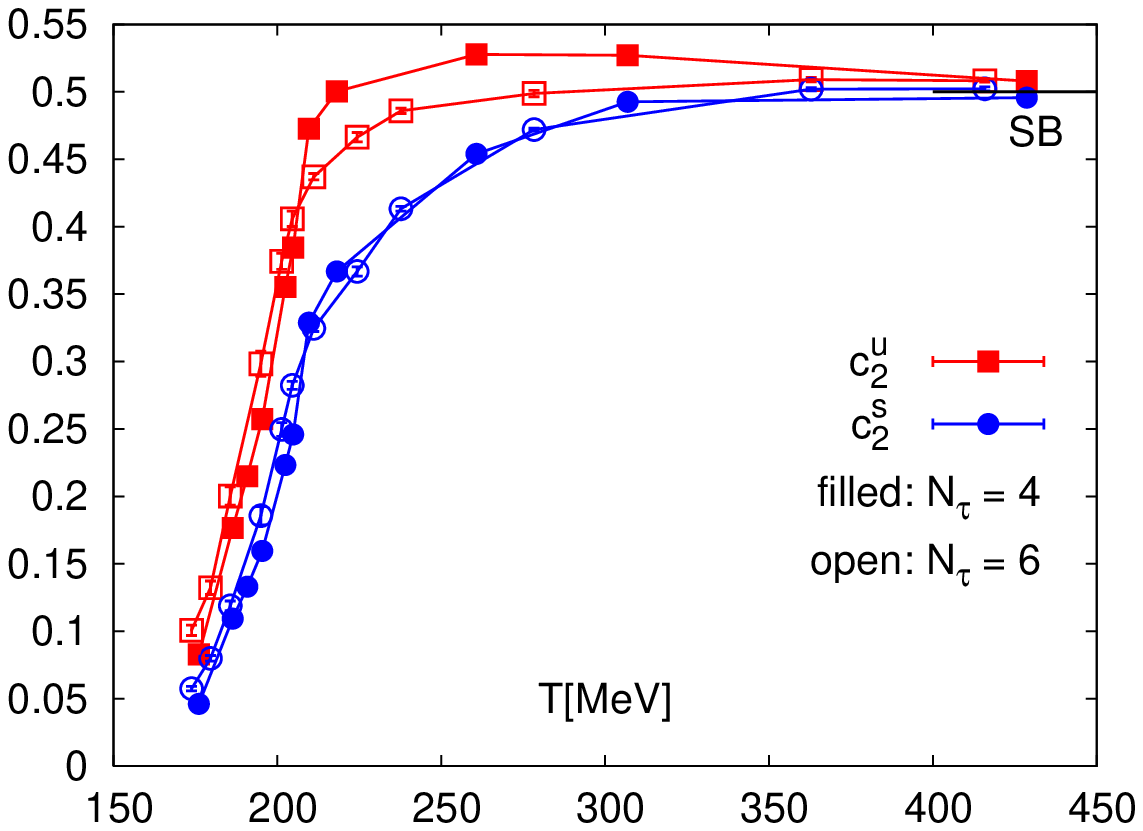}
\includegraphics[width=0.49\textwidth,height=0.2\textheight]{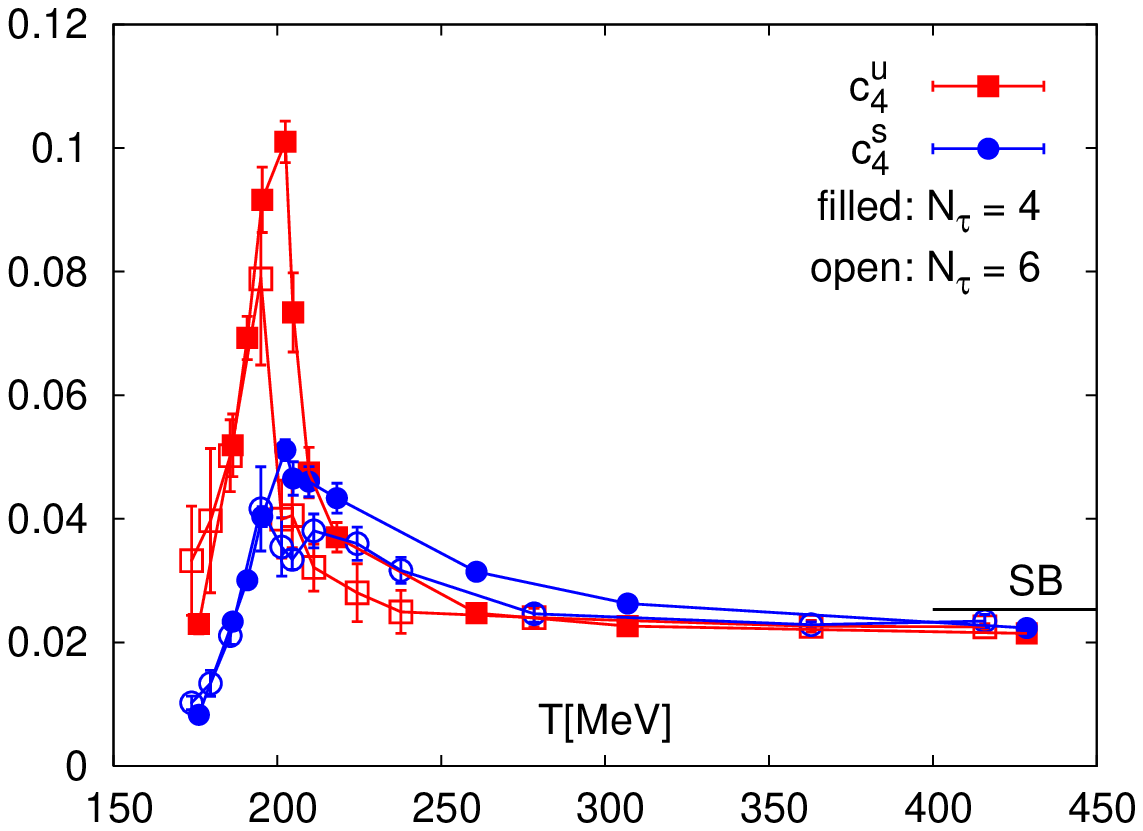}
\caption{Staggered quark results for the Taylor coefficients 
$c_2^{u,s}$ and $c_4^{u,s}$, obtained with (2+1)-flavor of p4fat3 
fermions on $N_\tau=4$ and $6$ lattices~\cite{cschmidt_and_cmiao}.}
\label{fig:c2c4_stag}
\end{figure}

%These generalized susceptibilities are interesting for a number of
%reasons. 
In Fig.~\ref{fig:c2c4_stag}, we show results for $c_2^X$ and $c_4^X$
for $X=u,~s$ calculated with staggered fermions \cite{cschmidt_and_cmiao}.
It is apparent that $c_2^X$ varies rapidly 
and $c_4^X$ peaks within a small temperature range. This behavior
may be interpreted as signaling a transition from a confined phase
consisting of heavy hadrons to a deconfined phase consisting of
partons. In fact, the ratio $\chi_4^{(X)}/\chi_2^{(X)}$ is directly related 
to the squared unit charge of quantum number $X$, for the relevant degrees 
of freedom \cite{Ejiri:2005wq}.  

The phase diagram of the QGP at $\mu=0$ is by now believed to be understood
quite well. {\it E.g.}, we have plenty of evidence that the  
transition is a sharp crossover, rather than a true phase transition.
Hence, the general behavior of $c_2$ and $c_4$ as shown in 
Fig.~\ref{fig:c2c4_stag} will not change, even in the thermodynamic limit.
They may, however be sensitive to the values of the light quark masses.
At vanishing light quark mass, the transition at $\mu=0$ is expected to be 
second order, in the same universality class as 
the three-dimensional $O(4)$-models. 
In this case the appropriate scaling
field $t$ (reduced temperature) in the vicinity of the phase 
transition~\cite{hatta_ikeda}
depends on a combination of $T$ and $\mu$
\begin{equation} t = \left| \frac{T-T_c}{T_c}\right| +
A\left(\frac{\mu}{T_c}\right)^2.
\label{eq:scaling_field}
\end{equation} 
This explains the general structure of the temperature dependence 
of $c_{2n}$. We thus may expect that the
behavior of the $(2n)$-th $\mu$-derivative of the partition function is
similar to that of the $n$-th temperature derivative;
e.g. $\epsilon \sim \partial \ln Z/\partial T\; \Leftrightarrow
\; c_2 \sim \partial^2 \ln Z/\partial \mu^2$.
%(see Table~\ref{tab:T-mu_analogy}). 
A plot of $c_2^X$ versus the
temperature thus shall resemble the corresponding plot for the energy
density, while that of $c_4^X$ shall resemble the specific heat $C_v$.

We expect this general pattern of the expansion coefficients also to
be reproduced in calculations with domain wall fermions.
Details of the temperature dependence, in particular at low temperature, 
may, however, be 
sensitive to chiral properties of the lattice discretization scheme used.
Our main interest in performing these
calculations also with domain wall fermions thus is to understand
to what extent the better chiral symmetry of this fermion action shows
up in thermodynamic observables that are directly sensitive to the
hadron spectrum.

\subsection{Beyond the Staggered Formulation}
Chiral symmetry plays an important role in determining the nature of
the QGP transition. In the staggered formulation, the full $SU(3)_L
\times SU(3)_R$ chiral symmetry is explicitly broken; at non-zero
values of the lattice spacing only two  $U(1)$ subgroups
remain; one in the light and strange quark sector, respectively. 
As a consequence universal properties at the chiral
phase transition (in the chiral limit) may be recovered only
in the continuum limit. Moreover, the 
loss of full chiral symmetry leads to too few light degrees of freedom
at low temperatures~{\it e.g.} one light pion instead of
three, which may influence bulk thermodynamic properties in this
regime. Furthermore, questions have been raised about the validity of
``the rooting trick'' used for staggered fermions~\cite{creutz}, which 
at present are not fully settled~\cite{creutzxx,kronfeld}.

It is clearly desirable to go beyond the staggered formulation and
work with a fully chiral formulation. Domain Wall Fermions (DWF) meet 
these
criteria~\cite{shamir}. In this formulation, one introduces a fifth
dimension $(x,y,z,t,s)$, so that the lattice dimensions are
$N_\sigma^3 \times N_\tau \times L_s$. Only fermions can propagate
along the fifth direction; the gauge fields exist only on the
4-dimensional slices $s=0,1,\dots L_s-1$. For this reason, the Domain
Wall formulation can also be thought of as a theory of $L_s$ fermion
flavors coupled in a nontrivial way. When one simulates the theory,
one finds, among other massive modes, two solutions that satisfy~\cite{fleming}   
\begin{equation} 
\left[M_{dwf}(p)\right]_{s,s'}\  \Psi^{(\pm)}_{s'}\ (p) = \left( -i \sum_{\mu=1}^4 \sin^2 p_\mu \right) \Psi^{(\pm)}_s\ (p),
\end{equation} 
where $M_{dwf}$ is the Domain-Wall matrix and a sum over $s'$ is implied. Furthermore, these modes are eigenstates of $\gamma_5$ and their wavefunctions are found to be 
localized on the two hyper-planes $s = 0$ and $s = L_s-1$
respectively. These are exactly the chiral modes we want. Using a  
{\it Pauli-Villars subtraction} allows to remove contributions of
the heavy modes~\cite{vranas} which leaves us with a theory that is,
for $L_s\rightarrow \infty$, chiral-invariant under the full symmetry group.
At finite values of $L_s$ a so-called residual mass leads to 
violations of chiral symmetry. In practice one is forced to perform
calculations at small lattice spacings to suppress these residual 
mass effects. For thermodynamics calculations this means that
calculations should be performed for  $N_\tau \ge 8$ and/or sufficiently
large $L_s$ to suppress the residual mass effects. In fact, the 
calculations we present here only marginally satisfy these 
constraints \cite{mcheng} and thus should be considered as a first
feasibility study.  

\section{Details of the Simulation}
We use here finite temperature 2+1-flavor dynamical domain-wall 
fermion ensembles generated by the RBC-Collaboration ~\cite{mcheng}. 
The lattice dimensions are $16^3\times8\times32$. The light
and strange quark masses used in these calculations are 
$am_q = 0.003$ and $am_s = 0.037$, and the
domain-wall height and anisotropy factor were $aM_5 = 1.8$ and $a/a_5 =
1.00$ respectively.  The gauge configurations have been generated
at several different values of the coupling, ranging from $\beta=1.95$ 
to $\beta=2.14$. This covers the transition region from low to
high temperatures. On these data sets we started to calculate
 the lowest order expansion coefficients. 

The expansion coefficients can entirely be expressed in terms of 
traces of the (5-dimensional) DWF fermion matrices over space-time, 
color and spin indices. For the 
diagonal and off-diagonal coefficients we obtain from derivatives of the 
fermion matrices for light ($M_u$) and strange ($M_s$) quarks, 
\begin{equation} c_2^X =
\frac{1}{2VT}\left\{\Bigg\langle\tr\left(\Minv\ddM{X}\right)\Bigg\rangle
- \Bigg\langle\tr\left(\Minv\dM{X}\Minv\dM{X}\right) \Bigg\rangle +
\Bigg\langle \tr^2 \left(\Minv\dM{X}\right) \Bigg\rangle \right\}
\label{eq:c2_as_traces}
\end{equation}
and
\begin{equation}
 c_{11}^{XY} = \frac{1}{VT}
\Bigg\langle\tr\left(\Minv\dM{X}\right)
\tr\left(\Minv\dM{Y}\right)\Bigg\rangle,
\label{eq:c11_as_traces}
\end{equation}
respectively. Here we used the shorthand notation $M\equiv M_X$ for $X=u,\; s$.
We estimated the traces using the stochastic random noise method with 
around $100-150$ random vectors for each trace. Products of traces
were evaluated in an unbiased manner. Some simulation details and our current
statistics are summarized in Table~\ref{tab:sim_details}.

\begin{table}[!tbh]
\label{tab:sim_details}
\begin{center}
\begin{tabular}{|r@{.}l|c|c|c|}
\hline
\multicolumn{2}{|c|}{$\beta$} & \# conf. & separation & \# r.v. \\
\hline
2 & 14 & 40 & 10 & 100 \\
2 & 11 & 35 & 10 & 100 \\
2 & 0625 & 111 & 10 & 150 \\
2 & 05 & 81 & 25 & 150 \\
2 & 0375 & 96 & 20 & 150 \\
2 & 025 & 71 & 20 & 150 \\
2 & 0125 & 125 & 10 & 150 \\
1 & 975 & 61 & 10 & 150 \\
1 & 95 & 73 & 10 & 150 \\
\hline
\end{tabular}
\end{center}
\caption{Details of the calculation: The columns give from left to
right the values of the lattice coupling $\beta$, the number of
evaluated configurations, the number of trajectories by which these
configurations are separated and the number of random vectors used for
the evaluation of the traces. }
\end{table}

\section{Results}
A word about the systematic errors: It can be shown that domain wall
fermions satisfy the same dispersion relation as naive
fermions~\cite{aarts_foley}. This implies that thermodynamic
quantities computed with the standard DWF action shall
have $\mathcal{O}(a^2)$ cut-off errors that are of the same 
magnitude as those of a standard staggered or naive fermion 
discretization scheme. This translates to $\sim 10\%$ cut-off errors at $N_\tau =
8$~\cite{karsch_laermann_scheredin_hegde}. Remarkably, the above
continues to be true even after the introduction of a chemical
potential, with only the magnitude of the correction
changing~\cite{karsch_laermann_scheredin_hegde,banerjee_gavai_sharma}. 
We note in passing that these $\mu$-dependent cut-off corrections can be 
computed exactly in the ideal gas
limit; they are given by Bernoulli polynomials.

\begin{figure}[!tbh]
\begin{center}
\includegraphics[width=0.49\textwidth]{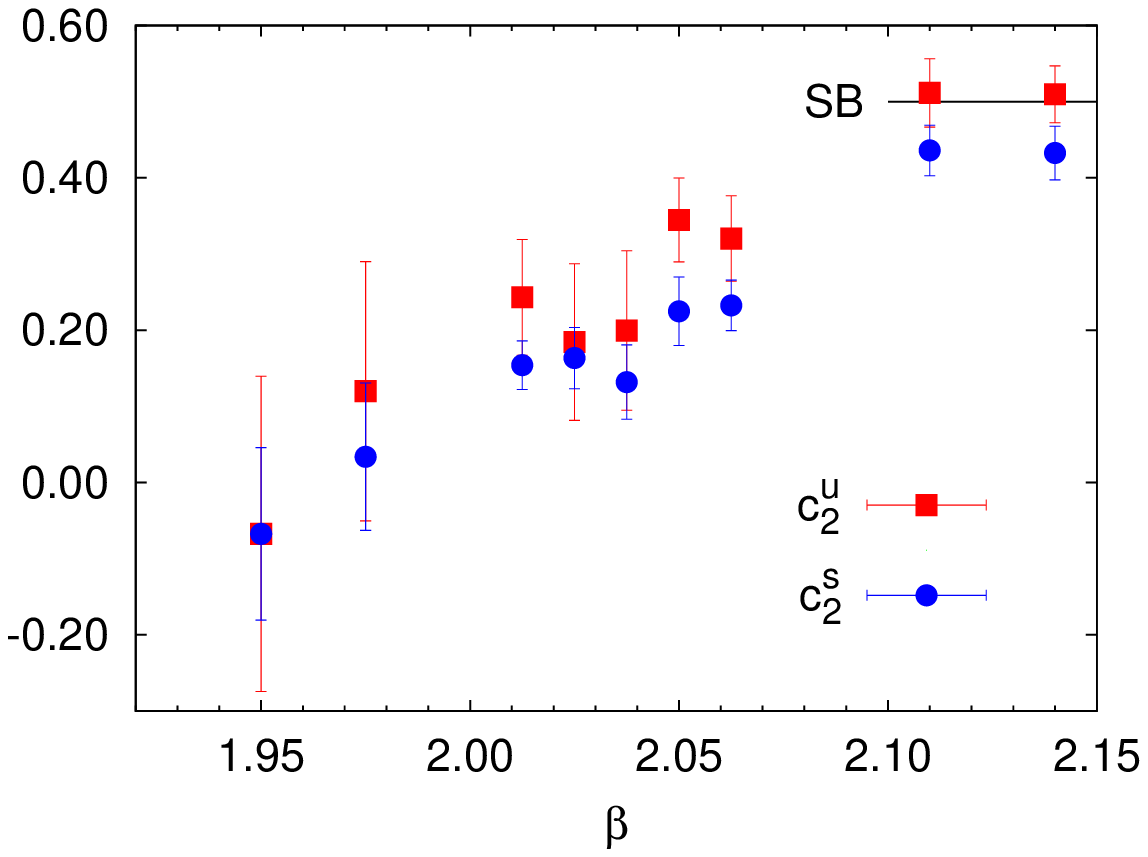} 
\includegraphics[width=0.49\textwidth]{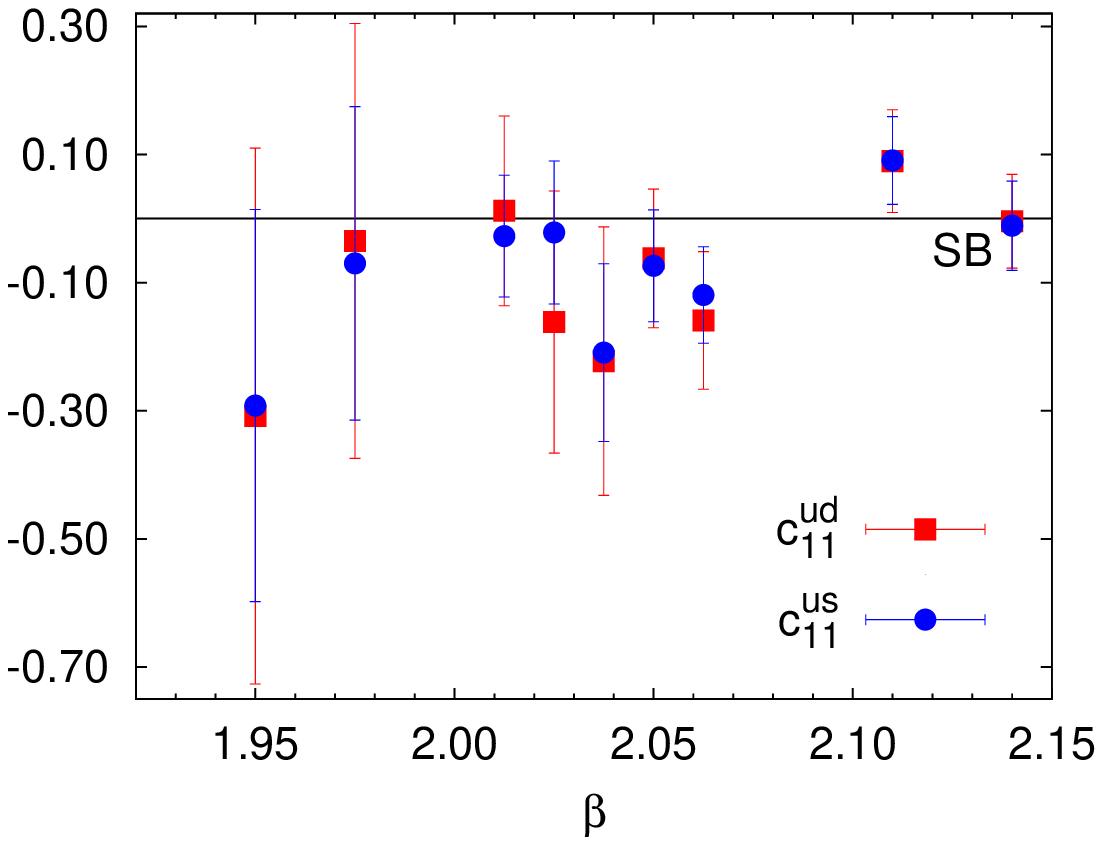}
\end{center}
\vspace{-0.4cm}
\caption{The expansion coefficients $\cu$, $\cs$ (left) and $\cud$,
$\cus$ (right) as a function of the coupling $\beta$.}
\label{fig:coff_quark}
\end{figure}

The statistical error in the susceptibilities was determined by the
jackknife method. There are two independent contributions to the
error: The error due to the finite size of the ensemble and the error
in the trace due to the finite number of random vectors. The latter
was found to be especially severe for the ``disconnected''
contributions, which are the products of traces in
Eqs.~\eqref{eq:c2_as_traces} and \eqref{eq:c11_as_traces}. 
This shows up in the large error bars for
the coefficients $\cu$, $\cs$ and especially the off-diagonal
coefficients $\cud$ and $\cus$, as can be seen from
Fig.~\ref{fig:coff_quark}. 
Although $\cu$ and $\cs$ do transit from a 
low value to a high one, it is difficult to assign a corresponding transition 
temperature or equivalent a coupling $\beta = \beta_d$. 

\begin{figure}[b]
\label{fig:results-2}
\centering
\includegraphics[width=0.49\textwidth]{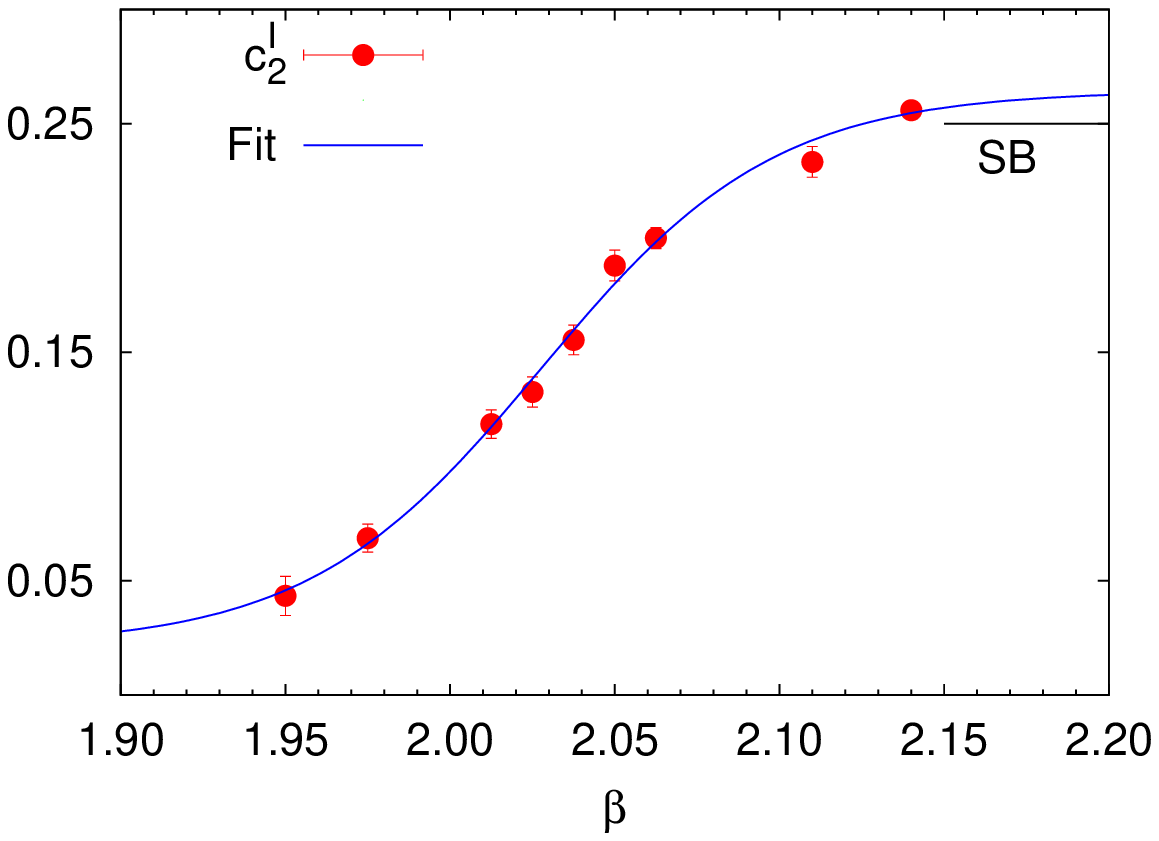} 
\includegraphics[width=0.49\textwidth]{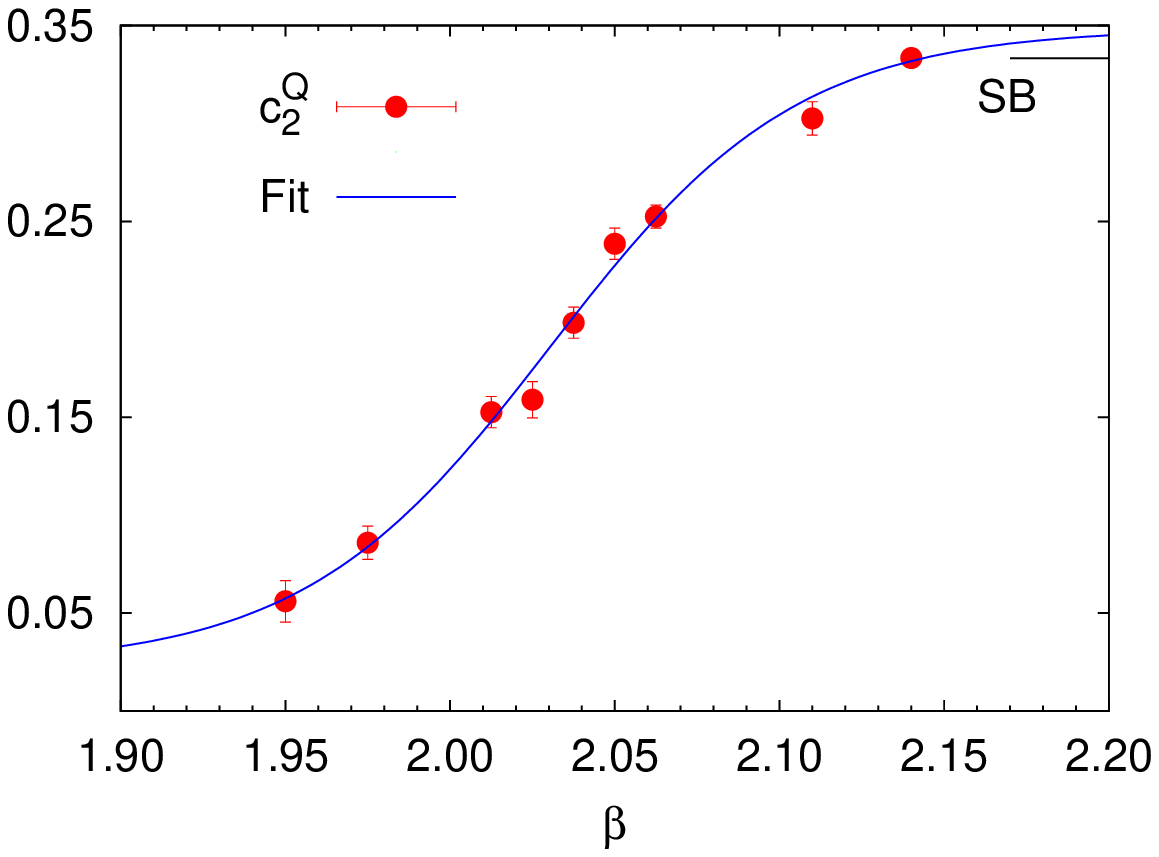}
%\vspace{-0.1\textheight}
\caption{$\cI$ and $\cQ$ as function of the lattice coupling $\beta$. 
The smooth curve is a fit of the ansatz given in
Eq.~\protect\ref{eq:ansatz} to the data; the best fit values are given in 
Table~\protect\ref{tab:best_fit_values}.}
\end{figure}

The disconnected contributions completely or
partially cancel each other in the iso-spin and electric charge
expansion coefficients $\cI$ and $\cQ$. This results in much smaller
errors for these two quantities as can be seen in
Fig.~\ref{fig:results-2}.
The coefficients $\cI$ and $\cQ$ show a smooth transition from a low
to a high value. There is some indication that they will overshoot the
Stefan-Boltzmann (SB) ideal gas values at higher temperatures. This
would be expected from the analysis of cut-off effects in thermodynamics
within the DWF formulation \cite{karsch_laermann_scheredin_hegde}. This 
deserves further analysis. 

As we do, at present, have no results on higher order expansion 
coefficients, e.g. $c_4^X$, which could directly be used to identify
a pseudo-critical coupling from the location of a peak in $c_4^X$, we
estimate a pseudo-critical coupling,  $\beta=\beta_d$, from fits
of $c_2^{(Q,I)}$. We use the ansatz~\cite{fleming}
\begin{equation} 
c_2^{(Q,I)} = A \tanh \left\{B(\beta-\beta_d)\right\} + C.
\label{eq:ansatz}
\end{equation}
The resulting curves are also shown in Fig.~\ref{fig:results-2},
superimposed on the data points. Table~\ref{tab:best_fit_values}
contains the best fit values of the free parameters $A$, $B$, $C$ and
$\beta_d$. 

For the transition value of the coupling we obtain $\beta_d
= 2.028(6)$ ($\beta_d = 2.030(7)$) from the fit to $\cI$ ($\cQ$). This value is in
agreement with the value $\beta_c = 2.031(5)$ obtained by
the RBC-collaboration~\cite{mcheng} from an analysis of the 
chiral condensate. We thus find that the deconfinement transition and chiral symmetry 
restoring transition occur in the same narrow temperature interval. 

From a determination of the Sommer parameter, $r_0 a^{-1}|_{\beta = \beta_d} = 3.25(18)$, the group was also able to deduce $T_c = 171(10)(17)$ MeV~\cite{mcheng}. While the latter error arose from the chiral and continuum extrapolations, the former was due to the fact that the residual mass varied across the temperature range and was significantly large for $\beta \lesssim \beta_c$.

We have stressed earlier that the current analysis is exploratory and
that, in particular, the residual mass effects inherent in the DWF
ansatz need to be reduced through calculations performed for large
values of $L_s$ and/or $N_\tau$. Clearly, this requires a more thorough 
analysis in the future.

\begin{table}[t]
\label{tab:best_fit_values}
\begin{center}
\begin{tabular}{|c|c|c|c|c|c|} \hline & $A$ & $B$ & $C$ & $\beta_d$ &
$\chi^2$/dof \\ 
\hline 
$\cI$ & 0.49(5) & 14.1(2.3) & 0.57(3) & {\bf
2.028(6)} & 1.028 \\ 
\hline 
$\cQ$ & 0.16(2) & 13.3(2.7) & 0.19(1) &
{\bf 2.030(7)} & 1.418 \\ 
\hline
\end{tabular}
\end{center}
\caption{The best fit values and their errors for the fit of 
$c_2^{(I,Q)}$ to the ansatz given in Eq.~\protect\ref{eq:ansatz}.
}
\end{table}

\section*{Acknowledgments} 
We thank Chulwoo Jung and
Michael Cheng for helpful discussions. The numerical simulations were
performed on the BlueGene/L computer at the New York Center for
Computational Science (NYCCS). The work of FK and PH has been
supported under Contract No. DE-AC02-98CH10886 of the U.S. Department
of Energy.

%%%%%%%%%%%%%%%%%%%%%%%%%%%%%%%%%%%%%%%%%%%%%%%%%%%%%%%%%%%%%%%%%%%%%%%%%%%%%%%%%%%%%%%%%%%%%%%%%%%%%%%%%%%%%%%%%%%%%%%%%

%%%%%%%%%%%%%%%%%%%%%%%%%%%%%%%%%%%%%%%%%%%%%%%%%%%%%%%%%%%%%%%%%%%%%%%%%%%%%%%%%%%%%%%%%%%%%%%%%%%%%%%%%%%%%%%%%%%%%%%%%

\end{document}